\documentclass[pre,showpacs,preprint,superscriptaddress]{revtex4}

\usepackage{hyperref}
\usepackage{graphicx}
\usepackage{color}
\usepackage{bm,amsmath}

\newcommand{\bsf}[1]{\textsf{\textbf{#1}}}
\newcommand{\crrp}{c(\mathbf{r}-\mathbf{r}')}
\newcommand{\lsim}{\:\raisebox{-0.5ex}{$\stackrel{\textstyle<}{\sim}$}\:}
\begin{document}

\title{The mode-coupling glass transition in a fluid confined by a periodic potential}

\author{Saroj Kumar Nandi}
\email{snandi@physics.iisc.ernet.in}
\affiliation{Centre for Condensed Matter Theory, Department of Physics, Indian Institute of Science, Bangalore 560 012, India.}

\author{Sarika Maitra Bhattacharyya}
\email{mb.sarika@ncl.res.in}
\affiliation{Complex Fluids \& Polymer Engineering Group, National Chemical Laboratory, Dr. Homi Bhabha Road, Pune  411 008, India.}

\author{Sriram Ramaswamy}
\email{sriram@physics.iisc.ernet.in}
\altaffiliation[Also at ]{JNCASR, Bangalore 560 064, India.}
\affiliation{Centre for Condensed Matter Theory, Department of Physics, Indian Institute of Science, Bangalore  560 012, India}

\begin{abstract}
We show that a fluid under strong spatially periodic confinement displays a glass transition within mode-coupling theory (MCT) at a much lower density than the corresponding bulk system. We use fluctuating hydrodynamics, with confinement imposed through a periodic potential  whose wavelength plays an important role in our treatment. To make the calculation tractable we implement a detailed calculation in one dimension. Although we do not expect simple $1d$ fluids to show a glass transition, our results are indicative of the behaviour expected in higher dimensions. In a certain region of parameter space we observe a three-step relaxation reported recently in computer simulations [S.H. Krishnan, PhD thesis, Indian Institute of Science (2005); Kim \textit{et al.}, Eur. Phys. J-ST \textbf{189}, 135-139 (2010)] and a glass-glass transition. 
We compare our results to those of Krakoviack, PRE \textbf{75} (2007) 031503 and Lang \textit{et al.}, PRL \textbf{105} (2010) 125701. 
\end{abstract}

\pacs{64.70.P-, 64.70.Q-}
\maketitle

\section{Introduction and results}
\label{intro}

Confined between two atomically smooth surfaces a few molecular diameters apart, fluids depart markedly from their bulk behaviour, displaying a prodigious increase in viscosity and structural relaxation times, and shear-thinning and viscoelasticity at remarkably low shear-rates and frequencies. The consensus from experiments \cite{demirel96,granick91,klein95} in a Surface Force Apparatus adapted for shear studies \cite{israelachvili88} and simulations \cite{gao97,ayappa07,kob02,kob04,ghatak01,thompson90,thompson95,cui01,cui03} is that strong  confinement moves the system into the regime of the glass transition. Such a trend is indeed found theoretically \cite{krakoviack07,krakoviack09} by adapting mode-coupling theory (MCT) \cite{goetzebook,reichman05,spdas04} to the case of a fluid confined in a porous medium modelled as a random aggregate of hard spheres. MCT for a fluid between two smooth planar walls shows similar slowing down \cite{lang10}. Strictly planar walls, however, exert no force parallel to themselves, so that the dynamics remains momentum-conserving and long-wavelength density fluctuations travel as sound waves in directions parallel to the walls, unless no-slip is imposed by hand. Questions about the nature of finite-size scaling  \cite{smarajit09,biroli06} at the glass transition, and the related issue of cooperatively rearranging regions \cite{kob02}, provide further motivation for studying the effect of confinement on the slowing down of the dynamics of a liquid. The extraction of viscosity and friction parameters from confined fluid flow experiments is discussed in \cite{persson94,saroj06}. 

In this paper we study the dynamics of a confined dense fluid, in a coarse-grained approach. Confinement enters the theory through an external potential; the resulting static inhomogeneous density background is a surrogate for the potential in our calculation. The mean density and temperature of the system are encoded in the static structure factor of the fluid in the absence of confinement. Our work differs in detail from those of Refs. \cite{krakoviack07,lang10} in our use of the fluctuating hydrodynamic approach (see, e.g., \cite{spdas04}), encoding the interactions between particles and confining medium in the free-energy functional, and examining the problem in detail in one space dimension. 

\begin{figure}
\includegraphics[width=8.2cm]{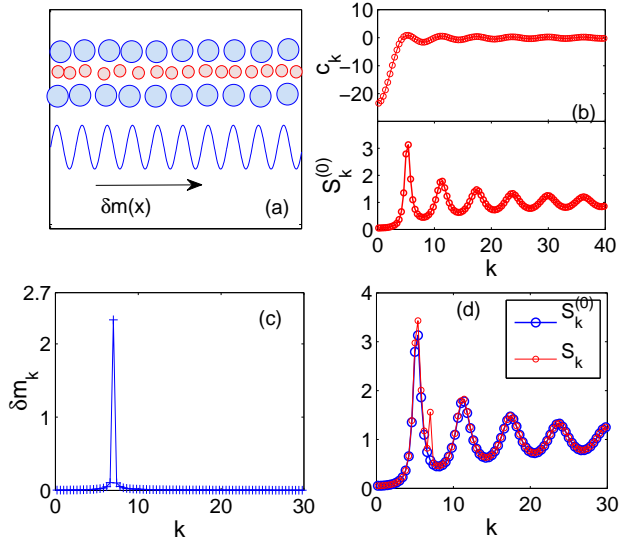}
\caption{(Color online) (a) A schematic presentation of a one-dimensional fluid confined by periodic external walls, which we re-express in terms of a periodic mean density profile. (b) The direct correlation function $c_k$ and the static structure factor $S_k^{(0)}$ for a one-dimensional bulk fluid. (c) The Fourier transformed background density $\delta m_k$. (d) The bulk structure factor $S_k^{(0)}$ and its modification $S_k$ due to the external potential.}
\label{charac}
\end{figure}

We summarize our main results below and in Figs. \ref{charac} - \ref{threestep}. (i) We present a particularly transparent  derivation, from the equations of fluctuating hydrodynamics, of the mode-coupling equations for the memory function and time-correlation function of the density field of a confined fluid. (ii) In the fluid phase, confinement renders the density dynamics diffusive at long wavelengths, with a diffusivity calculable, via mode-coupling, in terms of properties of the inhomogeneous background density field. (iii) We show in detail (see Fig. \ref{changingpot}) that strong enough confinement can drive the system through a glass transition, in conditions under which the system in bulk would be a fluid. To make the calculation tractable we work in one dimension, where a glass transition is unlikely for a fluid without confinement. However the structure of the calculation makes it clear that similar behaviour is expected in realistic higher-dimensional systems. (iv) The strength of potential required to produce the transition is lowest when the wavelength $\ell$ of the potential matches the length scale corresponding to the structure factor peak of the fluid (Fig. \ref{phasediagram}). (v) For densities $\rho_0$ below a value $\rho_c$ which depends on $\ell$, a continuous onset of the glassy state is observed as the potential strength is increased (Fig. \ref{continuous}). (vi) In the glassy state for $\rho_0 \lsim \rho_c$, we predict a three-step relaxation of the density correlation function (Fig. \ref{threestep}), as seen in recent molecular dynamics studies \cite{krishnan05,kuni_epjst2010} of confined fluids. (vii) Correspondingly, for a certain range of densities, two thresholds of confinement strength are seen for the onset of the non-ergodicity parameter: first a continuous onset, then a discontinuous jump. 

The prediction of a transition in one dimension, the finding that a periodic background suffices to enhance the transition, the emergence of diffusive dynamics for the density, and the multistep relaxation of the collective density correlator in the continuous transition regime are all features that distinguish our work from existing theoretical studies \cite{krakoviack07,lang10} of the glass transition under confinement. In particular, our approach offers a natural MCT-based explanation of the multiple plateaux in the intermediate scattering function seen in the simulations of \cite{krishnan05,kuni_epjst2010}. The double onset of the non-ergodicity parameter seen in our calculations provides a much stronger example of the confinement-induced glass-glass transition than that seen in \cite{krakoviack07}, and is similar to the behaviour found in \cite{krakoviack09} for the tagged-particle density, for confinement in a disordered medium.

The rest of this paper is organized as follows. Section \ref{confinedmctsection} begins with a discussion of possible dynamical effects of confinement in section \ref{genremarks}, and continues in section \ref{gencalc} with general conclusions from MCT for the fluctuating hydrodynamics of confined fluids. Section \ref{detailedcalc} presents our calculation in detail for a model one-dimensional fluid, leading to a phase diagram as a function of density and confinement strength. We close in section \ref{disc} with a discussion and summary.  

\section{Mode-coupling theory for a fluid in a confining potential}
\label{confinedmctsection}

\subsection{Remarks on the dynamics of confined fluids}
\label{genremarks}
Before entering into the calculations that led to the above results, some general remarks are necessary. The viscosity measured in the surface force apparatus (SFA) experiments corresponds to gradients along the $z$ direction normal to the plates and velocity in the $xy =\, \perp$ plane. The density modulations contributing to these components of the viscosity must have a wavevector component in the $\perp$ plane, because shear with gradient along $z$ does not directly affect structures varying only along $z$. In fact, the relevant fourier modes of the density must have wavevectors with nonzero $z$ \textit{and} $\perp$ components. Layering alone, with wavevector only along $\hat{\mathbf{z}}$, will not suffice. Next, what features of confinement are responsible for the dramatic slowing down? It seems reasonably clear that the pressure applied in the SFA experiments is not directly responsible, as it does not amount to a bulk hydrostatic compression, but simply goes into determining the thickness of the confined fluid layer. The confining walls rule out motion normal to their surface; they limit motion parallel to their surface through no-slip; and they alter the static structure of the fluid through steric or potential interactions. Can no-slip alone produce a significant slowing down of the density? To make this question concrete, consider a fluid with viscosity $\eta$ and velocity field $\mathbf{v}$, with density field $\rho$ governed by a free-energy functional $F[\rho]$ (for example, a Ramakrishnan-Yussouff \cite{ramakrishnan79} functional), lying between a pair of walls parallel to the plane, separated by distance $w$ along the $z$ direction. If we impose no-slip at the walls, the dynamics of the in-plane velocity field, on length scales in the $\perp$ plane large compared to $w$ is governed by lubrication: $(\eta / w^2) \mathbf{v}_{\perp} \simeq -\rho \nabla_{\perp} \delta F / \delta \rho$. The continuity equation $\partial_t \rho = - \nabla \cdot (\rho \mathbf{v})$ then implies an effective two-dimensional dynamics 
\begin{equation}
\label{densitydiff}
\partial_t \rho = \nabla_{\perp} \cdot \left({w^2 \rho \over \eta} \rho \nabla_{\perp} {\delta F_{eff} \over \delta \rho} \right) + \nabla_{\perp} \cdot \mathbf{f}_{\perp}.
\end{equation}
In writing (\ref{densitydiff}) we have implicitly averaged over $z$, so that $\rho$ depends only on $x$ and $y$, 
and we have written the $z$-averaged force density on the right-hand side in a form similar to the bulk, with an effective free-energy density $F_{eff}$.  
Eq. (\ref{densitydiff}) is of the type analysed in \cite{miyazaki_reichman,biman}, and $\mathbf{f}$ is a multiplicative noise whose form \cite{dean,miyazaki_reichman,abhiksr} guarantees that equal-time correlations of the two-dimensional density field are governed by $F_{eff}$.  To linear order in density fluctuations, at long wavelength, (\ref{densitydiff}) is a diffusion equation, with diffusivity $k_BTw^2\eta^{-1}\rho_0(1-\rho_0c_{q=0})$, where $c_q$ is the direct pair correlation function of the confined fluid, and $k_B$ and $T$, respectively, are Boltzmann's constant and absolute temperature. Nonlinear effects enter via the interactions embodied in $F_{eff}$, and can lead to a slowing down with increasing density and decreasing temperature \cite{miyazaki_reichman,kirk_thiru}.
The confinement scale $w$ in (\ref{densitydiff}) appears in two different ways. One, explicitly, in determining the diffusivity $\sim w^2/\eta$, and two, implicitly, within $F_{eff}$. 
The vanishing of the bare diffusivity $\propto w^2$, a consequence of no-slip alone, can readily be absorbed into a rescaling of the time in (\ref{densitydiff}), and cannot be further enhanced in an MCT feedback mechanism \cite{fuchspc}. A decrease in $w$ can lead to a glass transition only through structural inputs via $F_{eff}$. We are thus obliged to consider the effect of the walls on the structure of the fluid and thence on its dynamics \cite{notekob}.

\subsection{Fluctuating hydrodynamics of confined fluids}
\label{gencalc}

As we remarked in the Introduction, our treatment of the effect of confinement will replace walls by an external potential. Consider a fluid with velocity field $\mathbf{v}$ and density \cite{footnote2} field $\rho$ with mean $\rho_0$, in the presence of an externally imposed potential $U(\mathbf{x})$. The equations of fluctuating hydrodynamics for an isothermal fluid, extended down to the length-scales relevant to the slow dynamics of a fluid near structural arrest \cite{spdas04}, are the continuity equation 
\begin{equation}
\label{conteq}
\partial_t \rho+\nabla\cdot(\rho{\bf v})=0,
\end{equation}
and the generalized Navier-Stokes equation 
\begin{equation}
\label{NSeq}
\rho(\partial_t + \mathbf{v}\cdot\nabla)\mathbf{v}=\eta\nabla^2{\bf v}+(\zeta+\eta/3)\nabla\nabla\cdot{\bf v}-\rho\nabla\frac{\delta F^U}{\delta\rho} +\mathbf{f},
\end{equation}
where $\zeta$ and $\eta$ are the \textit{bare} bulk and shear viscosities, thermal fluctuations enter through the Gaussian white noise $\mathbf{f}$ with $\langle \mathbf{f}(\mathbf{0},0) \mathbf{f}(\mathbf{r},t)\rangle = -2k_BT [\eta \bsf{I}\nabla^2 + (\zeta + \eta/3)\nabla \nabla]\delta(\mathbf{r})\delta(t)$. 

Let us linearise the equation of motion by ignoring $\delta\rho {\bf v}$ in (\ref{conteq}) and (\ref{NSeq}), and replace velocity in the divergence of Eq. (\ref{NSeq}) using Eq. (\ref{conteq}) to obtain the equation of motion for the density fluctuation alone
\begin{equation}
\frac{\partial^2\delta\rho({\bf r},t)}{\partial t^2}=D_L\bigtriangledown^2\frac{\partial\delta\rho({\bf r},t)}{\partial t}+\nabla\cdot(\rho\nabla\frac{\delta F^U}{\delta\rho})+\nabla\cdot{\bf f},
\end{equation}
where $D_L=(\zeta+4\eta/3 )/\rho_0$. We take the space-Fourier transform of the above equation and obtain
\begin{equation}
\label{confineddensityeq}
\frac{\partial^2\delta\rho_{\bf k}(t)}{\partial t^2}+\Gamma_k\frac{\partial\delta\rho_{\bf k}(t)}{\partial t}=\bigg[\nabla\cdot(\rho\nabla\frac{\delta F^U}{\delta\rho})\bigg]_{k}-i{\bf k}\cdot{\bf f}_{\bf k}^L,
\end{equation}
where $\Gamma_k=D_Lk^2$ is the {\em bare} longitudinal damping coefficient.

The confining potential $U$ has been incorporated into the density-wave free-energy functional \cite{ramakrishnan79} 
\begin{align}
\label{modifiedRY}
\beta F^U&=\int_{\bf r}\left\{\rho({\bf r})\ln \frac{\rho({\bf r})}{\rho_0}-[\rho({\bf r})-\rho_0]\right\}\nonumber\\
-&\frac{1}{2}\int_{{\bf r},{\bf r}'}\crrp[\rho({\bf r})-\rho_0][\rho({\bf r}')-\rho_0]
+\beta\int_{\bf r}U({\bf r})\rho({\bf r}) 
\end{align}
where $\int_{\bf r}\equiv \int \text{d}{\bf r}$, $\beta = 1 / k_B T$, and $\crrp$, the direct pair correlation function \textit{in the absence of }$U$, is a coarse-grained expression of the intermolecular interactions in the fluid. Our reference state is $m(\mathbf{r})$, the inhomogeneous equilibrium density field in the presence of $U$, which satisfies $\delta F^U[m]/\delta m(\mathbf{r}) =0$, i.e., 
\begin{equation}
\label{staticdensity}
\ln \frac{m(\mathbf{r})}{\rho_0}= -\beta U({\bf r}) + \int_{{\bf r}'}\crrp \delta m({\bf r}') 
\end{equation}
where $\delta m({\bf r}) \equiv m({\bf r})- \rho_0$.
Writing $\rho({\bf r},t)= \rho_0 + \delta m(\mathbf{r}) + \delta\rho({\bf r},t)$, the force density from (\ref{modifiedRY}), after replacing $U({\bf r})$ in terms of $m({\bf r})$ using Eq. (\ref{staticdensity}), takes the form
\begin{align}
\label{forceeqn}
\rho\nabla \frac{\delta \beta F^U}{\delta \rho({\bf r},t)}&=\nabla\int_{{\bf r}'}[\delta({\bf r}-{\bf r}')-\rho_0\crrp]\delta\rho({\bf r}',t)\nonumber\\
&-\nabla\int_{{\bf r}'}\delta m({\bf r})\crrp\delta\rho({\bf r}',t)\nonumber\\
&-\frac{\nabla m({\bf r})}{m({\bf r})}\int_{{\bf r}'}[\delta({\bf r}-{\bf r}')-m({\bf r})c(\bf r-\bf r')]\delta\rho({\bf r}',t)\nonumber\\
&-\delta\rho({\bf r},t)\nabla\int_{{\bf r}'}\crrp\delta\rho({\bf r}',t). 
\end{align}
Note that the effect of confinement is contained only in the background density field $m(\mathbf{r})$; $U$ itself does not appear explicitly. The hydrodynamic equations (\ref{conteq}) and (\ref{NSeq}) with (\ref{forceeqn}) then readily yield the dynamical equation \cite{inertiafootnote} 
\begin{equation}
\label{eqofmotion1}
(\partial_t^2 + \Gamma_{\mathbf{k}} \partial_t +{k_B T k^2 \over S_{\mathbf{k}}^{(0)}})\delta\rho_{\bf k}=-i \mathbf{k}\cdot (\mathbf{F}_{\mathbf{k}}+\mathbf{f}_{\mathbf{k}}) \equiv -i \mathbf{k}\cdot (\mathbf{F}^{m\rho}_{\mathbf{k}} + \mathbf{F}^{\rho\rho}_{\mathbf{k}} + \mathbf{f}^L_{\mathbf{k}})
\end{equation}
for the spatial Fourier transform $\delta \rho_{\mathbf{k}}(t)$ of the density field, with force densities 
\begin{equation}
\label{Fmrho}
\mathbf{F}^{m\rho}_{\mathbf{k}}(t) = ik_BT \int_{\bf q}[\mathbf{k} c_{\mathbf{q}}+({\bf k}-{\bf q})/\rho_0 S^{(0)}_{\mathbf{q}}]\delta m_{{\bf k}-{\bf q}}\delta\rho_{\bf q}(t)
\end{equation}
arising from interaction with the static inhomogeneous background, to first order in $\delta m(\mathbf{r})$, which suffices for the one-loop treatment we will present \cite{footnote3}, and
\begin{equation}
\label{Frhorho}
\mathbf{F}^{\rho\rho}_{\mathbf{k}}(t) = \frac{i}{2} k_BT \int_{\bf q}[{\bf q}c_{\mathbf{q}}+({\bf k-\bf q})c_{\mathbf{k}-\mathbf{q}}]\delta\rho_{\bf q}(t)\delta\rho_{\bf k-q}(t) 
\end{equation}
from the pairwise interaction of fluid density fluctuations. In (\ref{eqofmotion1}) $\mathbf{f}^L_\mathbf{k}(t)$ is the longitudinal part of the Fourier-transform of the bare noise $\mathbf{f}$ in (\ref{NSeq}), and $S^{(0)}_{\mathbf{q}}$ in (\ref{eqofmotion1}) and (\ref{Fmrho}) is the static structure factor of the bulk fluid without confinement. In (\ref{Fmrho}) and (\ref{Frhorho}) and hereafter $\int_{\bf q}$ denotes $\int d^d q / (2 \pi)^d$. The autocorrelation of $\mathbf{f}^L_{\mathbf{k}}(t)$ is linked to the \textit{bare} longitudinal damping $\Gamma_{\mathbf{k}} \equiv D_L k^2$, and the Kubo formula \cite{kubo} tells us that the excess damping due to interactions, expressed in the time domain, is given by the memory function 
\begin{equation}
\label{renormfric}
\mathcal{M}_{\mathbf{k}}(t) = {1 \over k_B T V}\langle \mathbf{F}_{\mathbf{k}}(0) \cdot \mathbf{F}_{-\mathbf{k}}(t)\rangle
\end{equation}
where $V \to \infty$ is the system volume, and $\mathbf{F}_{\mathbf{k}}(t)$ is as in (\ref{eqofmotion1})-(\ref{Frhorho}). 

We define translation-invariant correlation functions $S_{\mathbf{k}}(t) \equiv \int d^dr \exp(-i \mathbf{k} \cdot \mathbf{r})[\langle \delta \rho (\mathbf{r}_0)(0) \delta \rho (\mathbf{r}_0 + \mathbf{r})(t)\rangle]_{\mathbf{r}_0}$ where $[\,\, ]_{\mathbf{r}_0}$ denotes an average over ${\mathbf{r}_0}$ over a period of the background. Within a Gaussian decoupling approximation (\ref{eqofmotion1})-(\ref{renormfric}) lead to \cite{kawasaki03} 
\begin{equation}
\label{eqofmotion2}
\ddot{\phi}_{\bf k}(t)+D_Lk^2\dot{\phi}_{\bf k}(t)+\frac{k_BTk^2}{S_k}\phi_{\bf k}(t)
+\int_0^t \mathcal{M}_k(t-\tau)\dot{\phi}_{\bf k}(\tau)d\tau=0
\end{equation}
for $\phi_{\mathbf{k}}(t) \equiv S_{\mathbf{k}}(t)/S_{\mathbf{k}}(0)$, with 
\begin{align}
\label{memory}
\mathcal{M}_{\mathbf{k}}(t)&=\frac{C_1}{2k^2}\int_{\bf q}[{\bf k}\cdot{\bf q}c_q+{\bf k}\cdot({\bf k}-{\bf q})c_{k-q}]^2 S_{{\bf k}-{\bf q}}(t)S_{\bf q}(t)\nonumber\\
+&\frac{C_1}{k^2}\int_{\bf q} [{\bf k}\cdot{\bf q}c_q+{\bf k}\cdot({\bf k}-{\bf q})/\rho_0]^2S_{{\bf k}-{\bf q}}^bS_{\bf q}(t), 
\end{align}
where $C_1={k_BT\rho_0}$ and $S_{\mathbf{k}}^b \equiv \int d^dr \exp(-i \mathbf{k} \cdot \mathbf{r})[\delta m (\mathbf{r}_0)\delta m (\mathbf{r}_0 + \mathbf{r})]_{\mathbf{r}_0}$ is the structure factor of the background density field. Eqs. (\ref{eqofmotion2}) and (\ref{memory}) are the equations of mode-coupling theory (MCT) for a confined fluid, as an initial value problem for $S_{\mathbf{k}}(t)$ given the static structure factor $S_{\mathbf{k}}\equiv S_{\mathbf{k}}(t=0)$ of the \textit{confined} fluid. These equations are identical to those obtained by Krakoviack \cite{krakoviack07} through the projection operator formalism for the case of a fluid confined by a porous medium. Our approach offers a simple and transparent derivation of these results.

Irrespective of the details of the confining medium, (\ref{memory}) leads us to our first result, an important general feature of confined MCT. Of the two contributions to $\mathcal{M}_{\mathbf{k}}(t)$ on the right-hand side of (\ref{memory}), the second, coming from the interaction of dynamic density fluctuations with the background density field, is non-vanishing for wavevector $\mathbf{k}\to 0$. The interaction of the fluid with inhomogeneities of the confining medium damps the flow even at zero wavenumber, and MCT provides a convenient way to calculate this effective ``Darcy'' damping. It is this contribution that leads to an effective no-slip condition on confining walls endowed with structure, periodic or otherwise. As a consequence, for long timescales and small $k$ the third and fourth terms on the left-hand side of (\ref{eqofmotion2}) dominate, reducing it, as is to be expected \cite{sriram82,hagenetal97}, to a diffusion equation for $\phi_{\mathbf{k}}(t)$ with collective diffusivity $k_B T /S_{\mathbf{k} = \mathbf{0}} \mathcal{M}_{\mathbf{0}0}$ where $\mathcal{M}_{\mathbf{0}0}$ is the Fourier-transform of the memory function at zero wavenumber and frequency. The crossover to Brownian rather than Newtonian dynamics assumed for convenience in \cite{krakoviack07,franosch97} \cite{footnote5} emerges at small wavenumber in a calculable form as a result of interaction with the confining medium. 

\section{Mode-coupling in one dimension in a periodic potential}
\label{detailedcalc}
The theory presented in the previous section is quite general and can, in principle, be applied in any dimension. The mode-coupling calculation for a system with spatially periodic confinement in two or three dimensions makes enormous demands on computational time, because of the loss of isotropy. We therefore chose to display the principle of the calculation by working in one dimension. We are aware that a simple fluid of, say, hard rods in one space dimension has no crystalline phase and therefore cannot be supercooled or overcompressed, rendering it a poor candidate for a glass transition. As MCT is relatively insensitive to spatial dimensionality, we expect nonetheless that the trends in our $1d$ MCT treatment will be found in calculations in dimensions $\geq 2$. We will return to the question of the behaviour of one-dimensional systems at the end of the paper.

In order to implement our calculation we need an expression for $S_{\mathbf{k}}$ for the confined fluid. Rather than taking it from experiment or liquid theory, we express it in terms of the structure factor $S^{(0)}_\mathbf{k}$ of the bulk fluid which we treat as given. We expand (\ref{modifiedRY}) around the static inhomogeneous density $m({\bf r})$; $S_{\mathbf{k}}$ is then determined by the coefficient of the term quadratic in $\delta \rho$. This crude approximation is adequate for the purpose of illustration, and can be improved upon if necessary. To linear order in $\delta m_{\mathbf{k}}$ it is straightforward to see that 
\begin{equation}
\label{skconf}
S_{\bf k}=S_k^{(0)}+\frac{1}{\rho_0}S_k^{(0)} \delta m_{\bf k}S_k^{(0)}.
\end{equation}

We must solve the closed set of equations (\ref{eqofmotion2}) - (\ref{skconf}) numerically to obtain the behaviour of our confined fluid. As remarked above, we do this for space dimension $d=1$. The calculation can be viewed as a schematic MCT for the confined fluid problem in which the structure of the confining walls is taken into account in a simple manner. 

\begin{figure}
\includegraphics[width=16.5cm]{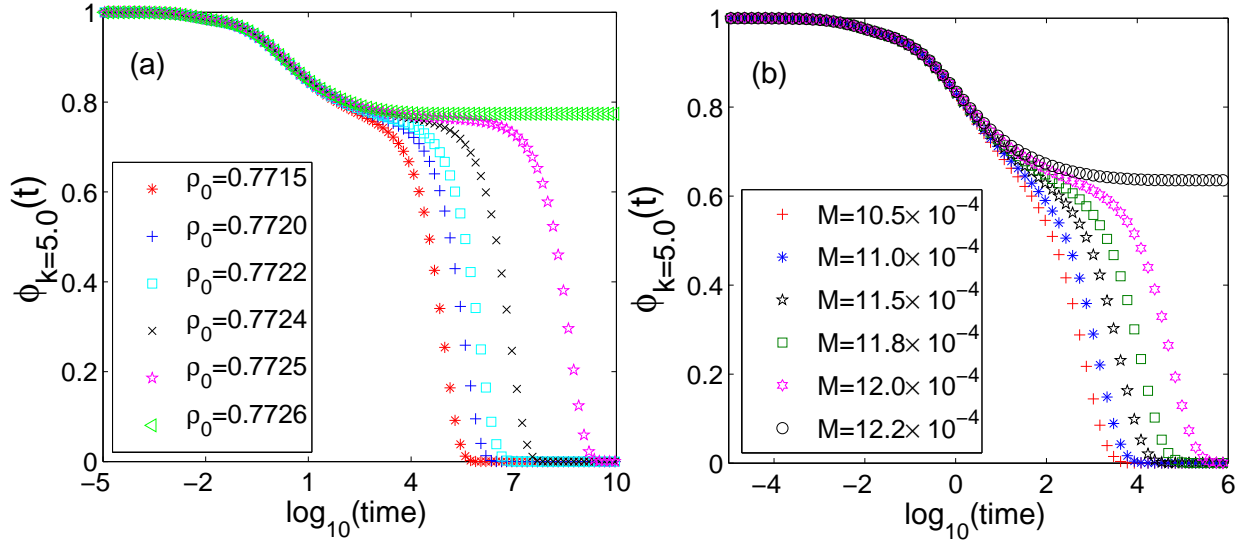}
\caption{(Color online) (a) The relaxation of the normalised coherent intermediate scattering function $\phi_{\bf k}(t)$ for $k=5.0$ for an unconfined one-dimensional fluid shows the mode coupling transition at density $\rho_0=0.7726$. (b) We set the density $\rho_0=0.70$ so that the system is in a fluid state [$\phi_{\bf k}(t)$ relaxes to zero] in the absence of an external potential. For an external potential with period $\ell=1.3$, corresponding to first maximum of the structure factor of the fluid without confinement, we increase $M$ from a small value and observe the MCT transition at $M=12.2 \times 10^{-4}$. The plot shows the relaxation of $\phi_{\bf k}(t)$ for $k=5.0$.}
\label{changingpot}
\end{figure}

We first solved the MCT equations (\ref{eqofmotion2}) and (\ref{memory}) \textit{without confinement}, in one dimension, using the direct correlation function for a one-dimensional hard-rod fluid \cite{wertheim64} as input, and found an MCT glass transition at $\rho_0=0.7726$ (Fig. \ref{changingpot} (a)). The nature of the MCT calculation in $d=1$ deserves some explanation. Through the quadratic nonlinearity, one mode at wavevector $\mathbf{k}$ couples to two others at $\mathbf{k}_1$, $\mathbf{k}_2$ with $\mathbf{k}_1 + \mathbf{k}_2 = \mathbf{k}$. For $d \geq 2$ all three of these modes can be taken to lie  on the first shell of maxima $|\mathbf{k}| = k_0$ of the structure factor $S(\mathbf{k})$. How then do we understand the MCT glass transition qualitatively for $d=1$, where the ``shell'' is two points? The answer: for a $1d$ fluid the higher-order peaks of $S(\mathbf{k})$ carry substantial weight; the dominant triple of modes coupled by MCT must consist of two with $|k| = k_0$ and one with $|k| = 2 k_0$. Having established an MCT glass transition in $d=1$, we now proceed to examine its modification by confinement. 

\begin{figure}
\includegraphics[width=16.2cm]{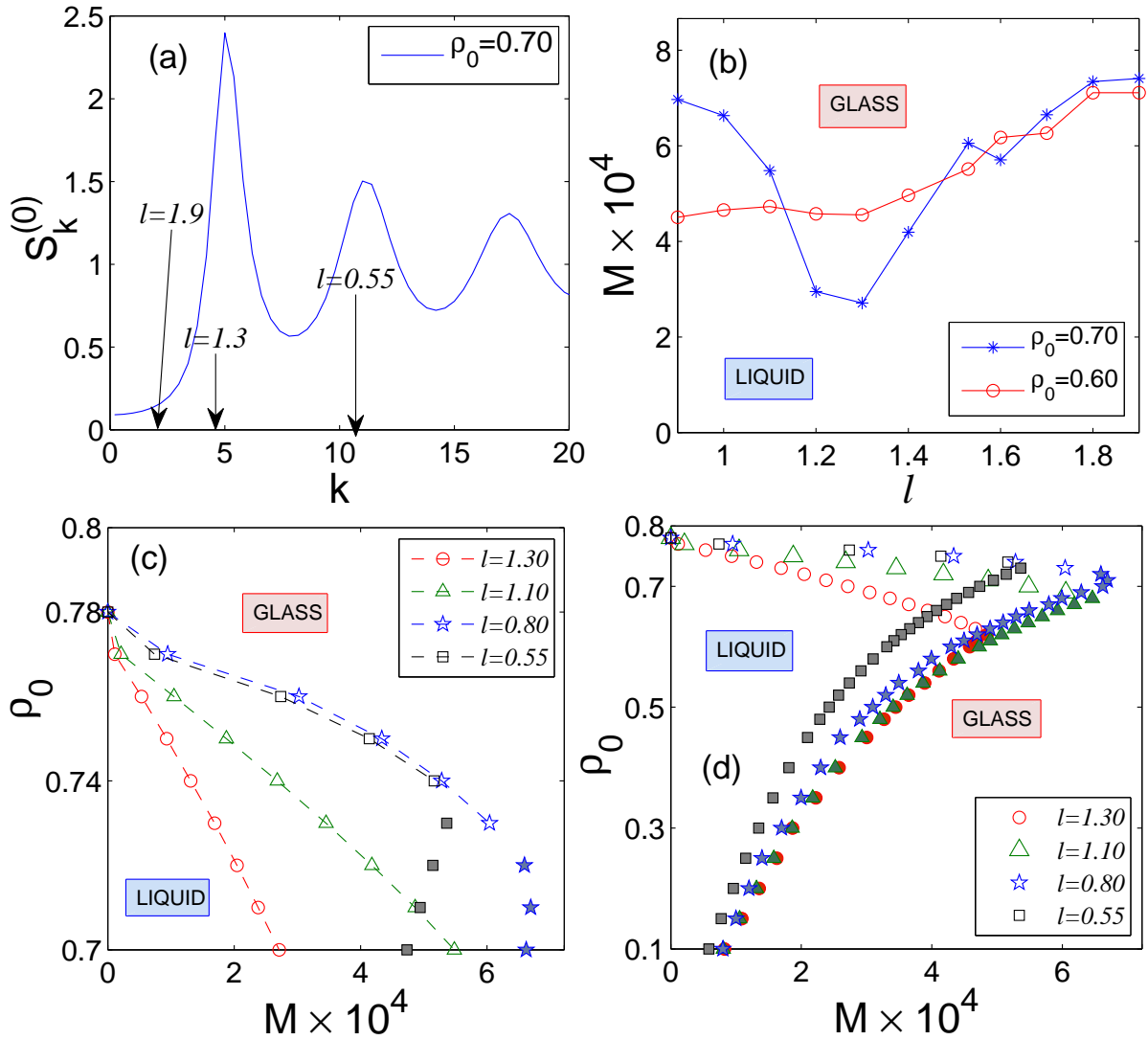}
\caption{(Color online) (a) The static structure factor $S_k^{(0)}$ for an unconfined  one-dimensional fluid at a density $\rho_0=0.70$. The various $l$ values marked on the $k$-axis in the figure correspond to the peak position of $\delta m_k$ [Fig. \ref{charac}(c)]. (b) The phase diagram for $M$ vs $\ell$ for two different densities. The threshold of $M$ is quite sensitive to $\ell$ for $\rho_0=0.70$ whereas it is less sensitive at $\rho_0=0.60$. (c) Phase diagram for $\rho_0$ vs $M$ for four values of $\ell$. (d) As in (c), for a larger range of $\rho_0$.  In both (c) and (d), filled and open symbols respectively indicate continuous and discontinuous transitions. Note: For clarity, the extension of the discontinuous transition line beyond the point where it crosses the continuous transition is not shown here [see Fig. \ref{threestep}}(a)].
\label{phasediagram}
\end{figure}

Consider one-dimensional confinement by a periodic potential $U(x)=U \sin {2\pi x \over \ell}$, 
in units of $k_B T$, where the parameters $U$ and $\ell$ control the strength and periodicity respectively of the potential. Given $u(x)$ we can construct the background density field $\delta m(x)=m(x)-\rho_0$ as in Eq. (\ref{staticdensity}). Instead of specifying $u(x)$, we will therefore characterize our confining medium by specifying $\delta m(x)$, which we take for simplicity to be sinusoidal, $\delta m(x) \simeq M \sin 2 \pi x/ \ell$. We set the density $\rho_0=0.70$, for which the system in the absence of a confining potential is in the fluid state, and take $\ell=1.3$, which means the period of the external potential is $2 \pi$ over the wavenumber at which the structure factor of the bulk one-dimensional fluid has its primary peak. As $M$ is increased past a threshold of about $12.2\times 10^{-4}$ we find, solving (\ref{eqofmotion2}) and (\ref{memory}) with (\ref{skconf}), that the mean relaxation time obtained from $\phi_{\mathbf{k}}(t)$ diverges, and the non-ergodicity parameter $f_k \equiv \lim_{t \to \infty} \phi_{\mathbf{k}}(t)$ jumps to a nonzero value (Fig. \ref{changingpot} (b)). This is the $1d$ MCT glass transition induced by confinement, i.e., by a periodic potential. 
Keeping $\rho_0$ fixed, we ask how changing $\ell$ affects the threshold $M$ for the transition. 
In Fig. \ref{phasediagram} (b), we scan around the first peak of the structure factor and find that at high density, the threshold value of $M$ is the lowest, i.e., the strongest enhancement of the transition is achieved, at a value of $\ell=1.3$ corresponding to the peak. Values of $\ell$ on either side of $1.3$ require a larger threshold value of $M$. However, at low density, the threshold of $M$ is very insensitive to $\ell$. This can be understood as follows. At higher densities, within MCT, the particles form a cage and $\ell=1.3$, being compatible with the preferred interparticle distance, facilitates this caging. For other values of $\ell$ the two length scales are different and, hence, $\delta m_k$ is less effective. At low density, there is no caging, so varying $\ell$ has little effect. It is important to note that the transition at $\rho_0=0.60$ is always continuous whereas at $\rho_0=0.70$ it can be continuous or discontinuous depending on the value of $\ell$ (Fig. \ref{phasediagram} (d)).

\begin{figure}
\includegraphics[width=8.2cm]{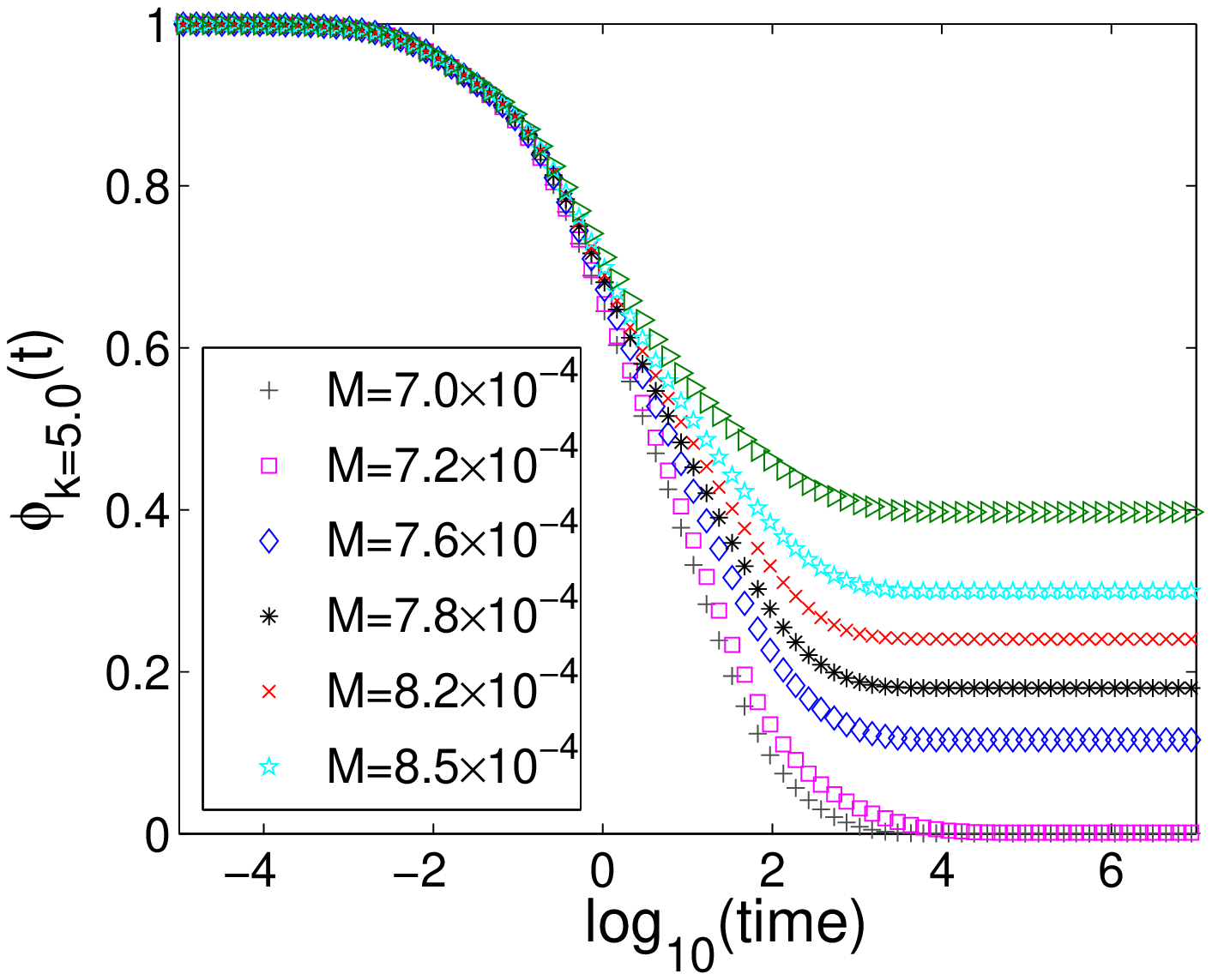}
\caption{(color online) At density $\rho_0=0.60$ and $\ell=0.80$, the non-ergodicity parameter $f_k=\lim_{t\to\infty}\phi_{\bf k}(t)$  shows a continuous onset to the MCT glass transition with increasing strength of the external potential. The plot shows the relaxation of $\phi_{\bf k}(t)$ for $k=5.0$.}
\label{continuous}
\end{figure}

We list several features of our phase diagram that are consistent with the findings of \cite{krakoviack07} for fluids in disordered porous media. For moderately high density $\rho_0$, the effect of the external potential can be seen as enhancing the correlations already present in the fluid, i.e., promoting the effect of the quadratic term in the MCT equation (\ref{memory}). At substantially lower densities, where correlations in the bulk fluid are weak, it is still possible within MCT to get a transition to a non-ergodic state by increasing $M$, but now the dominant role is presumably played by the term linear in $S_{\mathbf{q}}(t)$ in (\ref{memory}). Correspondingly, for densities $\rho_0$ lower than a value $\rho_c$ which depends on $\ell$, the transition turns continuous (Fig. \ref{continuous}). Quantitatively, however, the medium has a weaker effect at high densities than in \cite{krakoviack07}. More puzzling, it is substantially more effective at low densities than in \cite{krakoviack07}. Re-entrance is another feature observed in common with \cite{krakoviack07}: at densities somewhat below $\rho_c$ the threshold value of $M$ decreases with decreasing density. It is surprising that we observe it in our simple one-dimensional periodic system, and we suggest a simple interpretation. At low densities, the glassy state, if it exists, presumably comes from a feedback-enhanced slowing down of single-particle crossings of barriers posed by the external potential. As density is increased, interparticle repulsion becomes more important, and particles typically occupy regions not near the minima of the external potential. Thus, repulsion lowers the effective kinetic barrier to motion. Possibly our explanation is related to that proposed in \cite{krakoviack07}, but this is unclear. 

On the continuous MCT-glass transition line, there are possible difficulties with infrared divergences in the mode-coupling integral (\ref{memory}) because we are working in one dimension. This issue arises in a pronounced manner for a single particle in a disordered medium \cite{schnyder10}, where there is no transition in $d=1$ because the particle is always localized. There are however important differences between our system and that of \cite{schnyder10}. Our model is probably free from the $k \to 0$ problem because the background medium in our case is periodic, not disordered, and lacks weight at $k=0$. We have nevertheless checked that varying the lowest wavenumber over a limited range (from 0.1 to 0.2) does not alter, to 0.1 \% accuracy, the MCT transition density as estimated by the onset of a time-persistent density correlator. 

\begin{figure}
\includegraphics[width=16.2cm]{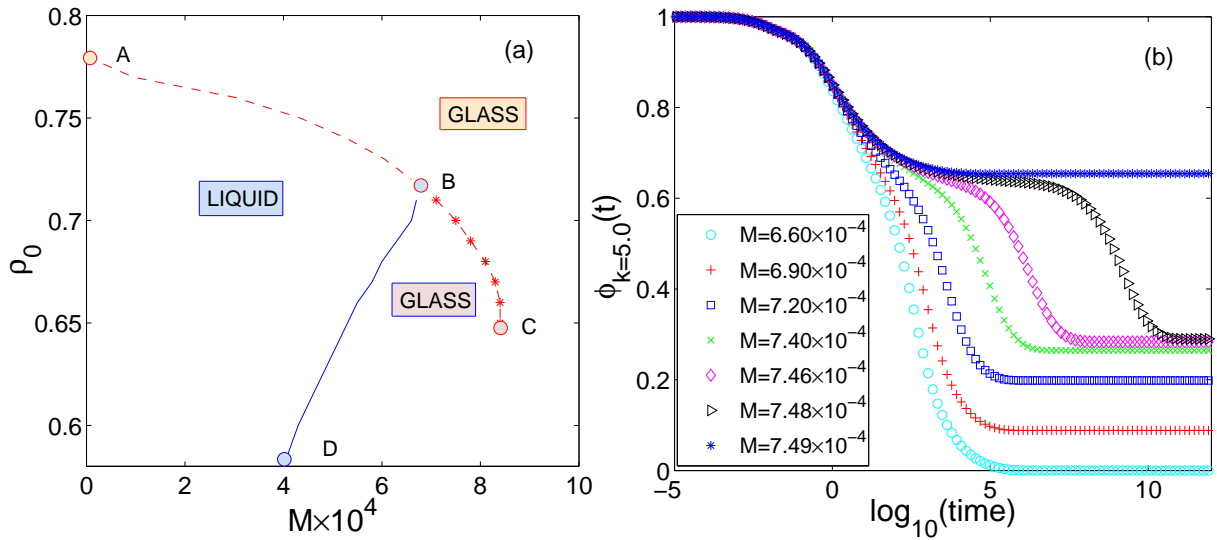}
\caption{(Color online) (a) For $0.65 < \rho_0 < 0.72$, $\ell = 0.8$, a continuous transition from liquid to glass (segment BD) is followed by a discontinuous glass-glass transition (segment BC). Note, however, that we do not resolve the phase boundary close to point B.
(b) If the density $\rho_0$ is close  to but lower than $\rho_c$, the threshold density where the transition switches from discontinuous to continuous, the density correlation function shows a three-step relaxation for strong enough background density. The figure shows the relaxation of $\phi_{k=5}(t)$ at a density $\rho_0=0.70$ as a function of $\log(\text{time})$ as we increase the strength of background density with $\ell=0.80$.}
\label{threestep}
\end{figure}

For $\rho_0$ moderately large, but still below $\rho_c$, $f_k$ undergoes two onsets as a function of $M$. Across a first threshold value of $M$, $f_k$ rises continuously from zero, presumably driven by the linear term in the MCT equation. Upon increasing $M$ past a second threshold, a discontinuous jump in $f_k$ is seen, which amounts to a glass-glass transition [Fig. \ref{threestep}(a)]. A hint of such behaviour has been reported in \cite{krakoviack07}; the effect appears far stronger in our case. The relaxation of the normalized intermediate scattering function in this second glassy state [see Fig. \ref{threestep}(b)] shows two plateaux followed by a third nonzero asymptotic value. Such dynamical behaviour has been reported for the self-intermediate scattering function in molecular dynamics studies of Lennard-Jones fluids under planar confinement to a thickness of about three molecules by structured walls \cite{krishnan05}, where it was called three-step relaxation. More recently similar dynamics has been reported for the coherent part of the intermediate scattering function for hard spheres in random media \cite{kuni_epjst2010}. The possibility of such multistep relaxation scenario was predicted in the context of various possible MCT integrals \cite{goetze88,fuchs91}, and obtained by Krakoviack \cite{krakoviack09} for tagged-particle motion, from MCT in a disordered medium. Such a complex relaxation scenario has also been reported in a variety of other contexts \cite{crisanti11,crisanti07,pinaki10}.

\section{Summary and Discussion}
\label{disc}
We have succeeded in providing an economical description of the slowing down of a fluid under confinement. We derived the mode-coupling equations through the dynamic density-wave approach \cite{footnote6} of fluctuating hydrodynamics, and implemented confinement in the form of an externally imposed periodic potential. In the spirit of a one-loop approach, we retained interactions between the fluid and the background density only to bilinear order, but this is not a serious difficulty; we could work with the background molecular field instead. In order to make our numerical calculations manageable we chose to work with a one-dimensional model and to deal with properties averaged over one period of the potential. Once we relax the period-averaging constraint, we can predict the degree of slowing down as a function of location in the potential. This is the nearest analogue, in our $1d$ model, to calculating properties as a function of distance from confining walls. Remarkably, our model calculation reproduces all the features observed in more detailed treatments, including a crossover to a continuous transition, the phenomenon of re-entrance, at lower densities, and a glass-glass transition as a function of confinement strength at intermediate densities. Unlike in \cite{krakoviack07}, quenched disorder plays no role, and the density dynamics is diffusive at small wavenumber, in contrast to \cite{lang10}. In future work we will consider in detail the problem of confinement in a planar geometry with structured walls, thereby improving on the treatments presented here. 

Apart from the glass-glass transition, which appears as a prominent feature in our treatment unlike in \cite{krakoviack07} where it is barely detectable, our most novel predictions are (i) that even a one-dimensional fluid, at least within an MCT approach, can undergo a glass transition; (ii) that an imposed periodic potential causes this transition to occur at lower densities and higher temperatures; (iii) that in a calculable range of densities, the relaxation of the density correlation function should take place through a three-step process, an effect for which there is evidence in molecular dynamics studies of two types of confined fluid systems \cite{krishnan05,kuni_epjst2010}, and as shown in MCT with confinement in a disordered medium \cite{krakoviack09} for the tagged-density correlation function. 

The initial experimental observations of the effect of confinement were done in shear flow measurements. It is therefore imperative to extend our approach to include shear. We have done this in a simplified implementation of the ideas of \cite{catesfuchs}. Our preliminary results \cite{sarojunpub} on an isotropized version of the calculation find that the shear-thinning moves to very low imposed flow-rates in the presence of the confining potential. Further results require imposing shear with planar confinement, preferably with structured walls. We are confident that our results will change only quantitatively when these elements of greater realism are introduced.

Lastly, should one expect a glass transition, or at least an MCT-style enhancement of viscosity, in experiments or simulations on one-dimensional fluids, confined or otherwise? It might seem unlikely, given that a $1d$ system has no crystalline phase past which one can supercool it to enter a region of metastability and glassiness. However, supercooling is not necessary for viscosity increase of the MCT variety \cite{taborek1986}, and numerical calorimetry \cite{kob09} finds precursors of glassy behaviour even for $1d$ Lennard-Jones mixtures. It might therefore be worth testing our ideas in simulations on $1d$ fluids, perhaps multicomponent or with complicated interactions, with and without confinement. In any case, we look forward to tests of our predictions in experiments and simulations, whether in one or higher dimensions.

\section{Acknowledgements} 
\label{ackno} 
We thank G. Ayappa, M. Cates, M. Das, C. Dasgupta, M. Fuchs, W. Kob, V. Krakoviack, and K. Miyazaki for discussions, and the DST, India, for support through a J C Bose Fellowship for SR. 

\bibliography{confinedfluid_final}

\begin{thebibliography}{10}
\bibitem{footnote2}%
  \BibitemOpen
  \bibinfo {note} {We set the masses of the constituent particles to unity and
  therefore do not distinguish between the mass density and number density}%
  \bibAnnoteFile{NoStop}{footnote2}%
\bibitem{footnote3}%
  \BibitemOpen
  \bibinfo {note} {We could instead directly work in terms of the potential $U$
  in which case the first order approximation will not be necessary. Our choice
  is in order to make contact with \cite{krakoviack07}}%
  \bibAnnoteFile{NoStop}{footnote3}%
\bibitem{footnote6}%
  \BibitemOpen
  \bibinfo {note} {The limitations of a continuum approach to these phenomena
  that take place at the scale of a few molecular layers are ultimately no
  worse than those of any continuum theory of the liquid-solid or liquid-glass
  transition}%
  \bibAnnoteFile{NoStop}{footnote6}%
\end{thebibliography}

\end{document}